# Fundamental guided electromagnetic dispersion characteristics in lossless dispersive metamaterial clad circular air hole waveguides

Ki Young Kim

*Department of Electrical Engineering and Computer Science, Northwestern University, Evanston, IL 60208, USA*

**Abstract**

The fundamental guided electromagnetic dispersion characteristics in lossless dispersive metamaterial clad circular air hole waveguides are investigated. Two operating guided modes are found to exist: circular waveguide and surface plasmon polariton modes that support fast and slow waves, respectively. Hybrid mode classifications are also made in an empirical manner so that the guided modes can be sorted into either TM-like or TE-like modes in terms of their unique dispersion characteristics. Unusual dispersion characteristics, including multi-valued propagation constants at a single frequency, backward waves, and subwavelength guided propagations, are observed and discussed in relation to the air-hole radii and dielectric and magnetic constants of the metamaterial clad. A discussion comparing the extraordinary dispersion characteristics with those of other dielectric, plasmonic, and metamaterial waveguides is also included.

## 1. Introduction

Metamaterials (MTMs) can be defined as artificial media whose electromagnetic properties are not usually observed in nature, since they simultaneously exhibit negative permittivity and permeability. Backward wave propagation (or left-handedness) and negative refraction are also representative fundamental properties of MTMs [1]. Since the initial experimental fabrication of artificial composite media with simultaneously negative permittivity and permeability [2], extensive research has been invested in seeking novel physical phenomena and innovative electromagnetic application areas, from radio to optical frequencies. As a result, many fascinating potential applications using MTMs have been released, some of which include microwave and optical cloaking devices [3, 4].

Studying extraordinary electromagnetic wave propagations along MTM waveguides is also a significant research field related to MTMs, as the wave guiding structure is fundamental and essential in various electromagnetic applications. The specific geometrical structures for MTM waveguides reported in previous literature include a single MTM slab [5-8], grounded MTM slab [9-12], multilayered or asymmetric MTM slab [13-16], air slab with MTM clad [17], MTM channel [18], MTM fiber [19-23], hollow fiber with MTM clad [24], plus many more. Some of the interesting novel phenomena that have rarely been encountered in conventional metallic or dielectric waveguides are backward waves [9, 24], the simultaneous propagation of forward and backward waves [8, 12, 24], superslow waves [12, 24], the suppression of specific guided modes [8, 10-13, 16], and power circulation [8, 9]. Thus, innovative optoelectronic functional devices utilizing one or several of these properties in MTM waveguides (see, *e.g.*, ref. [25]) are expected to be developed in the near future.

Accordingly, this paper investigates the guided dispersion characteristics of dispersive MTM clad circular air hole waveguides. In a previous study [24], the properties of surface plasmon modes were analyzed, focusing on the subwavelength guiding properties and power flows along the structure. However, to complete the dispersion analysis, this paper also includes circular waveguide modes that resemble those of conventional metallic circular waveguides in connection with surface plasmon modes. To obtain more general and deeper physical insights into this structure, various air hole radii are considered, along with systematic classifications of the guided modes, including hybrid modes. The physical meanings of the dispersion curves are also analyzed for several different air-hole radii. As for the MTM used in this study, a lossless dispersive MTM model is adopted, which has already been widely used to determine the guided dispersion characteristics of waveguiding structures with different geometrical configurations, thereby facilitating a comparative picture of MTM clad air circular waveguides with other geometry, which is one of the main interests of this study. A brief comparison of the guided electromagnetic dispersions with those for other dielectric, plasmonic, and MTM waveguides with similar and dissimilar geometrical configurations is also included.

## 2. Guiding structure and its associated characteristic equations

### *2.1. Guiding structure and material parameters*

Figure 1 shows a schematic view of the MTM clad circular air hole waveguide structure and associated cylindrical coordinates. The air hole region with radius $a$ is characterized by its unity values of dielectric and magnetic constants (relative permittivity and permeability), *i.e.*, $\varepsilon_{r1} = \mu_{r1} = 1.0$, where the subscript "1" denotes the air region ( $r < a$ ). This air hole is surrounded by a dispersive MTM, which is assumed to extend to infinity in the transverse direction and consist of the simplest MTM clad circular air hole wave guiding structure. The dielectric and magnetic constants of the dispersive MTM are given as follows

$$\varepsilon_{r2} = 1 - \left(\omega_p / \omega\right)^2 \qquad (1a)$$

$$\mu_{r2} = 1 - F\omega^2 / \left(\omega^2 - \omega_0^2\right). \qquad (1b)$$

where the subscripts "2" denote the MTM region ( $r > a$ ), $\omega\left(= 2\pi f\right)$ is the angular frequency, and the constants of $\omega_p / 2\pi = 10$ GHz , $\omega_0 / 2\pi = 4$ GHz, and $F = 0.56$ are given. This dispersive MTM model is motivated by experimental results [2, 26, 27] and has already been widely used in many theoretical MTM studies [5, 8, 9, 18-21, 24]. Figure 2 plots the dielectric and magnetic constants of the MTM model, where the dielectric and magnetic constants become negative below 10.0 GHz and 6.03 GHz, respectively. Thus, the double negative (DNG) permittivity and permeability region extends from 4.0 to 6.03 GHz, while 6.03 to 10 GHz belongs to the epsilon negative (ENG) region. Above 10.0 GHz, both material constants remain double positive (DPS) values. Points "*E*" and "*M*" in figure 2 are the frequency points where the dielectric and magnetic constants become negative unities, respectively, which play a significant role in the mode classification and dispersion characteristics of the present wave guiding structure. More discussion on this is given in section 3. Although another ENG region exists below 4.0 GHz (not shown in figure 2), this is outside the interest of this study and has been excluded.

### *2.2. Assignment of proper electromagnetic fields and characteristic equations for guided modes*

To derive a characteristic equation of the given structure, the electric and magnetic fields need to be defined in each region. Thus, to assign proper electric and magnetic fields in each region, allowed and forbidden regions for the guided modes should be considered and distinguished. Figure 3 shows the three possible cases for guided mode propagation in the present structure, in which $\bar{\beta}\left(=\beta/k_0\right)$ is the normalized propagation constant, *i.e.*, the propagation constant ($\beta$), which is normalized by the free space wave number ($k_0$). In the present study, $\mu_{r1}\varepsilon_{r1}=1.0$ is fixed for the air hole, yet $\mu_{r2}\varepsilon_{r2}$ can be variable with respect to the operating frequency due to the dispersive nature of each material constant, as given in (1).

For $\mu_{r2}\varepsilon_{r2}>0$ (either DPS or DNG), at first, the case of $\bar{\beta}>\left(\mu_{r2}\varepsilon_{r2}\right)^{1/2}>\left(\mu_{r1}\varepsilon_{r1}\right)^{1/2}$ in Fig. 3(a) or $\bar{\beta}>\left(\mu_{r1}\varepsilon_{r1}\right)^{1/2}>\left(\mu_{r2}\varepsilon_{r2}\right)^{1/2}$ in figure 3(b) can be considered. In either case, the transverse propagation constants in each region can be expressed as $k_1=k_0\left(\bar{\beta}^2-\mu_{r1}\varepsilon_{r1}\right)^{1/2}$ and $k_2=k_0\left(\bar{\beta}^2-\mu_{r2}\varepsilon_{r2}\right)^{1/2}$, respectively, allowing the axial electric and magnetic fields in each region to be expressed as follows according to the criteria in [28].

$$E_{z1}=AI_m\left(k_1 r\right)\exp\left\{j\left(\omega t-m\theta-\beta z\right)\right\} \qquad (2a)$$

$$H_{z1}=BI_m\left(k_1 r\right)\exp\left\{j\left(\omega t-m\theta-\beta z\right)\right\} \qquad (2b)$$

$$E_{z2} = CK_m\left(k_2 r\right)\exp\left\{j\left(\omega t - m\theta - \beta z\right)\right\} \tag{3a}$$

$$H_{z2} = DK_m\left(k_2 r\right)\exp\left\{j\left(\omega t - m\theta - \beta z\right)\right\} \tag{3b}$$

where *A*, *B*, *C*, and *D* are the real constants exhibiting the amplitudes of each field, $\exp\{\cdot\}$ is the propagation factor, *m* is the azimuthal eigen value, and $I_m(\cdot)$ and $K_m(\cdot)$ are a modified Bessel function of the first and second kind, respectively. Due to the properties of these functions with respect to the radial direction (*r*), the fields in the air and the MTM are decreased from the interface, *i.e.*, the fields are maximum at the interface exhibiting the typical property of a surface plasmon polariton (SPP) [29], so this mode is called the SPP mode and supports slow waves ($\beta > k_0$). Furthermore, due to the non-oscillatory nature of the fields in both regions, as sketched in figure 4(a), the SPP mode is only expected to have a principal mode, *i.e.*, exhibit a monomode property. This is obviously different from the cases of conventional optical fibers [28, 30] and circular metallic waveguides [31]. This circular SPP mode can also be found in a plasmonic fiber [32] (or plasma column [33]) and MTM fiber [19-23].

The next case is $\left(\mu_{r1}\varepsilon_{r1}\right)^{1/2} > \beta / k_0 > \left(\mu_{r2}\varepsilon_{r2}\right)^{1/2}$ in Fig. 3(b), in which the transverse propagation constants in each region can be expressed as $k_1 = k_0\left(\mu_{r1}\varepsilon_{r1} - \bar{\beta}^2\right)^{1/2}$ and $k_2 = k_0\left(\bar{\beta}^2 - \mu_{r2}\varepsilon_{r2}\right)^{1/2}$. The fields in the MTM region are evanescent, which is same as the previous case in order to fulfill the radiation condition at infinity, and the fields in the air core region are given as follows.

$$E_{z1} = AJ_m\left(k_1 r\right)\exp\left\{j\left(\omega t - m\theta - \beta z\right)\right\} \tag{4a}$$

$$H_{z1} = BJ_m\left(k_1 r\right)\exp\left\{j\left(\omega t - m\theta - \beta z\right)\right\} \tag{4b}$$

where $J_m(\cdot)$ is a Bessel function of the first kind. This case evidently belongs to the fast wave ($\beta / k_0 < \left(\mu_{r1}\varepsilon_{r1}\right)^{1/2}\left(=1.0\right)$) region and the fields in the air core region are oscillatory, as shown in figure 4(b), due to the property of $J_m(\cdot)$. Since this intensity profile in the air core region resembles those in a conventional circular waveguide [31], it is called the circular waveguide (CWG) mode. Following the standard steps of the boundary value problems from the axial field components in (2)-(4), the characteristic equation of the present wave guiding structure can be written as $F_1 F_2 = G^2$, where

$$F_1 = \frac{\varepsilon_{r1}}{k_1}\frac{I'_m(k_1 a)}{I_m(k_1 a)} - \frac{\varepsilon_{r2}}{k_2}\frac{K'_m(k_2 a)}{K_m(k_2 a)}, \qquad (5a)$$

$$F_2 = \frac{\mu_{r1}}{k_1}\frac{I'_m(k_1 a)}{I_m(k_1 a)} - \frac{\mu_{r2}}{k_2}\frac{K'_m(k_2 a)}{K_m(k_2 a)}, \qquad (5b)$$

$$G = \frac{m\beta}{k_0 a}\left(\frac{1}{k_1^2} - \frac{1}{k_2^2}\right) \qquad (5c)$$

for the SPP mode, and

$$F_1 = \frac{\varepsilon_{r1}}{k_1}\frac{J'_m(k_1 a)}{J_m(k_1 a)} + \frac{\varepsilon_{r2}}{k_2}\frac{K'_m(k_2 a)}{K_m(k_2 a)}, \qquad (6a)$$

$$F_2 = \frac{\mu_{r1}}{k_1}\frac{J'_m(k_1 a)}{J_m(k_1 a)} + \frac{\mu_{r2}}{k_2}\frac{K'_m(k_2 a)}{K_m(k_2 a)}, \qquad (6b)$$

$$G = \frac{m\beta}{k_0 a}\left(\frac{1}{k_1^2} + \frac{1}{k_2^2}\right) \qquad (6c)$$

for the CWG mode. Primes denote the differentiation with respect to the arguments.

Lastly, figure 3(c) is for the case of a single negative MTM clad, *i.e.*, $\varepsilon_{r2} < 0$ and $\mu_{r2} > 0$, or $\varepsilon_{r2} > 0$ and $\mu_{r2} < 0$. In either case, $k_2 = k_0\left(\bar{\beta}^2 - \mu_{r2}\varepsilon_{r2}\right)^{1/2}$ is always real and greater than zero, because $\mu_{r2}\varepsilon_{r2} < 0$. Thus, considering the sign of $\mu_{r2}\varepsilon_{r2}$ is out of the picture in the guided condition of figure 3(c), *i.e.*, $\mu_{r2}\varepsilon_{r2} < 0$ is a sufficient condition for the wave guiding, and the forbidden region, such as the dashed region in figures 3(a) and (b), has disappeared, the conditions of $\bar{\beta} > \left(\mu_{r1}\varepsilon_{r1}\right)^{1/2}$ and $\bar{\beta} < \left(\mu_{r1}\varepsilon_{r1}\right)^{1/2}$ correspond to the SPP and CWG modes, respectively, which are the same as the previous cases. This case corresponds to an ENG MTM clad, *i.e.*, $\varepsilon_{r2} < 0$ and $\mu_{r2} > 0(\neq 1.0)$, in which parts of the guided dispersion properties are expected to be similar to those of a plasmonic clad ( $\varepsilon_{r2} < 0$ and $\mu_{r2} = 1.0$ ) air hole waveguide [34]. Therefore, from the above discussions, the allowed and forbidden regions of the normalized propagation constants can be represented with respect to the operating frequency, as shown in figure 5, from which the normalized propagation constants obtained from the characteristic equation will be shown and their guided dispersion characteristics discussed with various air-hole radii, including subwavelength diameters.

## 3. Numerical results and discussion

### *3.1. TM-like modes*

For the circularly symmetric modes, *i.e.*, $m = 0$, the characteristic equation $F_1 F_2 = G^2$ can be separated into $F_1 = 0$ and $F_2 = 0$, corresponding to the $TM_{0n}$ and $TE_{0n}$ modes, respectively. Figure 6(a) shows the $TM_{01}$ mode for various radii of the air core. The $TE_{01}$ mode will be discussed in subsection 3.2. The dispersion curves for both the SPP and CWG modes seamlessly continued at their border ( $\beta = k_0$ ), even though they are governed by dissimilar characteristic equations of (5*a*) and (6*a*), respectively. When the radius of the hole was sufficiently large, *e.g.*, $a = 20\text{mm}$, the dispersion curve for the CWG mode was forward with a positive slope and cutoff at the line of $\bar{\beta} = \left(\mu_{r2}\varepsilon_{r2}\right)^{1/2}$. The powers carried along the air-core region and MTM region were positive and negative, respectively. Thus, the forward wave carried more power in the air region than in the MTM region. Meanwhile, this curve was deflected backward near 7.155 GHz in the SPP mode region. This curve with a negative slope approached 7.071 GHz backwardly, corresponding to frequency point "*E*" in figure 2, where the dielectric constant is negative unity. The cutoff frequency, *i.e.*, the frequency when $\bar{\beta} = \left(\mu_{r2}\varepsilon_{r2}\right)^{1/2}$ for the DNG region and $\bar{\beta} = 0$ for the ENG region, increased as the radius of the air core decreased. Backward waves in the CWG mode were found with air-hole radii smaller than $a = 14$ mm . The cutoff frequency in the CWG mode region shifted up to 10.0 GHz. This means that backward waves for the ENG clad air hole only existed for the frequency region corresponding to $-1.0 < \varepsilon_{r2} < 0$. Backward waves in the CWG mode also existed in the region of $\varepsilon_{r2} < -1.0$, yet this region was included in the DNG clad region, *i.e.*, below 6.03 GHz, as shown in the inset of figure 6(a). This may have been due to the DNG material effect in the given frequency range (below 6.03 GHz), as this kind of backward wave region has not been found in the ENG version of this waveguide, *i.e.*, a plasmonic clad circular air hole waveguide [34]. The band of the existing guided mode for $a = 14$ mm was very narrow and localized near frequency "*E*" (7.071 GHz). Waveguides with either larger or smaller radii than $a = 14$ mm had a broader guidance band (or existing band). When the radius of the air hole was large, a forward wave could exist, yet this disappeared when the radius of the air hole became small. Meanwhile, a small air core region directly indicated that the area over which the forward power carried became small.

Although higher order modes, such as $TM_{02}$, $TM_{03}$, … existed in the CWG mode region (not shown here) when the radius of the hole was sufficiently large, they were all suppressed when the air core hole was very small, *e.g.*, $a = 1$ mm. A more detailed discussion is not given, as the dispersion properties were quite similar to those of a plasmonic clad air hole waveguide [34] and the properties can easily be anticipated. Even when the refractive index of the air core was greater than that of the MTM clad in the DPS region above 10 GHz, *i.e.*, $(\mu_{r1}\varepsilon_{r1})^{1/2} > (\mu_{r2}\varepsilon_{r2})^{1/2}$ which is the same as the traditional index guiding condition [30], the principal modes did not exist.

If the azimuthal eigen value is not zero, the characteristic equation of $F_1 F_2 = G^2$ can no longer be clearly separated into the TM and TE modes and instead represents hybrid modes that have both TM and TE components. The characteristic equations for the hybrid modes can be rearranged as follows.

$$\left(\frac{\mu_{r2}}{\mu_{r1}} + \frac{\varepsilon_{r2}}{\varepsilon_{r1}}\right)\left(\frac{Q}{2}\right) \pm \sqrt{\left(\frac{\mu_{r2}}{\mu_{r1}} - \frac{\varepsilon_{r2}}{\varepsilon_{r1}}\right)^2 \left(\frac{Q}{2}\right)^2 + \frac{R^2}{\mu_{r1}\varepsilon_{r1}}} - P = 0\,, \tag{7}$$

where $$P_{CWG} = \frac{1}{k_1 a}\left(\frac{J_{m-1}(k_1 a)}{J_m(k_1 a)} - \frac{m}{k_1 a}\right), \tag{8a}$$

$$Q_{CWG} = \frac{1}{k_2 a}\left(\frac{K_{m-1}(k_2 a)}{K_m(k_2 a)} + \frac{m}{k_2 a}\right), \tag{8b}$$

$$R_{CWG} = \frac{m\overline{\beta}}{a^2}\left(\frac{1}{k_1^2} + \frac{1}{k_2^2}\right) \tag{8c}$$

for the CWG mode and

$$P_{SPP} = \frac{1}{k_1 a}\left(\frac{I_{m-1}(k_1 a)}{I_m(k_1 a)} - \frac{m}{k_1 a}\right), \tag{9a}$$

$$Q_{SPP} = -\frac{1}{k_2 a}\left(\frac{K_{m-1}(k_2 a)}{K_m(k_2 a)} + \frac{m}{k_2 a}\right), \tag{9b}$$

$$R_{SPP} = \frac{m\overline{\beta}}{a^2}\left(\frac{1}{k_1^2} - \frac{1}{k_2^2}\right) \tag{9c}$$

for the SPP mode.

Although complex mathematical evidence exists that can distinguish the HE and EH hybrid modes, this study discerned the hybrid modes empirically. Since the HE mode is referred to as a TM-like mode in a conventional circular dielectric rod waveguide [28], this convention was also followed in this study. The curves of the normalized propagation constant with a "–" sign in front of the square root in (7) approached point "*E*" when they were higher values, as shown in figures 6(b) and (c). Therefore, this mode was called the $HE_{m1}$ modes. Meanwhile, the curves with a "+" sign were assigned to the $EH_{m1}$ mode and their properties are discussed in the next subsection.

Since the higher order modes ($TM_{0n}$ and $HE_{mn}$ modes, $n \geq 2$) all belonged to the CWG mode (not shown here) and their dispersion characteristics can be easily anticipated based on the results of a plasmonic clad air hole waveguide [34], this study concentrated on the properties of the principal modes. The normalized propagation constants were mainly obtained in the ENG region (6.03 to 10 GHz), except when the radius of the hole was large. The normalized propagation constants of the CWG and SPP modes also continued seamlessly. The cutoff frequency (the frequency where $\beta / k_0 = 0$) shifted toward a higher frequency region as the radius of the hole decreased. As shown in Fig. 6, the dispersion characteristics between the symmetric mode ($TM_{01}$ mode) and the hybrid modes ($HE_{11}$ and $HE_{21}$ modes) were quite different.

Figure 7 shows the dispersion curves of the TM-likes modes in the case of $a = 15$ mm and $a = 1$ mm for the purpose of comparison. When the radius was large, an $HE_{11}$ single mode operation region existed, as in the case of conventional dielectric fibers. However, when the radius became small, the cutoff frequency of the $TM_{01}$ modes shifted and a single mode operation region of the $TM_{01}$ mode was generated. The single mode frequency of the $TM_{01}$ modes extended to the corresponding frequency band where the dielectric constant was –1.0 to 0 when the radius was very small. In the case of a plasmonic fiber or plasma column, this frequency is forbidden for the guided mode. (The guided modes of a plasmonic fiber are limited below the frequency of $\varepsilon_r < -1.0$ [32, 33].)

### *3.2. TE-like mode*

Figure 8 shows the TE-like mode dispersion characteristics of the MTM clad circular air hole waveguide. As in the previous case of the TM-like modes, only the principal modes for the three lowest order of azimuthal eigen values ($m = 0,\ 1,\ 2$) were considered. The normalized propagation constants obtained from the characteristic equation were all found in the DNG region. The normalized propagation constants of the $TE_{01}$ modes were obtained from (5*b*) and (6*b*), while those for the nonzero value of the azimuthal elgen value for the TE-like modes, *i.e.*, the EH modes, were obtained from (7) with a "+" sign. When the normalized propagation constants were higher, they approached point "*M*" in figure 2, *i.e.*, $f = 4.714$ GHz, corresponding to $\mu_{r2} = -1.0$.

In the case of the $TE_{01}$ mode, as shown in figure 8(a), the dispersion curves were all backward wave types for the given radii, and the normalized propagation constants became higher as the radius decreased, meaning that more power was carried in the MTM region. For the hybrid modes, the cutoff frequency (the frequency where $\beta / k_0 = \left(\mu_{r2}\varepsilon_{r2}\right)^{1/2}$) decreased when the radius of the hole decreased. Under certain circumstances, three guided modes in a single frequency were possible. For example, in the case of $a = 2mm$ in figure 8(b), there were two backward waves (negative slope) and one forward wave (positive slope). The dispersion characteristics of the $EH_{21}$ modes were quite similar to those of the $EH_{11}$ modes. The $EH_{21}$ modes also had forward waves when $a = 1, 2, 4$ mm in figure 8(c). As in the previous case of the TM-like modes, the dispersion properties between the circularly symmetric mode ($TE_{01}$ mode) and the hybrid modes ($EH_{11}$ and $EH_{21}$ modes) were quite different.

In the case of a plasmonic clad circular air hole waveguide, the TE-like mode only exists in the CWG mode region [34], however, both the SPP mode and the CWG mode were able to propagate along the MTM clad circular air hole waveguide, apparently due to the negative permeability.

Figure 9 shows the dispersion curves of the TE-like modes when $a = 10$ mm and $a = 1$ mm for the purpose of comparison with different radii. The cutoff frequencies of the hybrid modes shifted toward a lower frequency, while the normalized propagation constants became higher as the radii decreased. Thus, a slower backward $TE_{01}$ wave in a wider frequency range can be supported when the radii are smaller, which can be desirable when this structure is applied to optical devices using the backward wave property, *e.g.*, a contra-directional coupler [35].

Figure 10 shows the dispersion curves of the TM-like and TE-like modes when the radius of the air hole was $a = 1$ mm, which was quite small when compared with the operational wavelengths. The operation frequency bands of the TM-like and TE-like modes were well separated, the hybrid modes localized at frequencies corresponding to $\varepsilon_{r2} = -1.0$ and $\mu_{r2} = -1.0$ for the $HE_{m1}$ and $EH_{m1}$ modes, respectively, and the higher order modes ($n \geq 2$) in the CWG modes well suppressed. Therefore, the guiding structure could be simultaneously operated with different modes ($TM_{01}$ and $TE_{01}$ modes) in different frequency bands. This was absolutely due to the properties of the MTMs adopted in this study, *i.e.*, different frequency ranges for negative and dispersive parameters. Notwithstanding, in the near future, it will be possible to artificially design MTMs [36, 37] on the molecular level [38-40] with the desired operation region for certain optoelectronic devices.

Finally, the MTM used in this study was an ideal lossless model. Thus, in certain situations, such as a backward wave supporting case, where more power propagates along the MTM region, the dissipative loss might have a serious affect on the guided dispersion characteristics, which has not been considered here. However, recent studies have shown that a gain medium can compensate for dissipative losses in the SPP waveguides, enabling a substantial extension of the propagation length [41-43], which may be a possible solution to this problem.

## 4. Conclusions

The guided dispersion characteristics along frequency dispersive MTM clad circular air hole waveguides with various hole radii were investigated, and characteristic equations governing the dispersion characteristics obtained for SPP and CWG modes from proper field assignments in each guiding region. Hybrid modes were classified in an empirical manner so that all the guided modes could be classified as TM-like and TE-like modes. Detailed guided dispersion characteristics were discussed based on a well-known frequency dispersive MTM model using various air-hole radii. The subwavelength guidance of the electromagnetic waves along this waveguide was also discussed.

## Acknowledgments

This work was supported by Korea Research Foundation Grant funded by the Korean Government (MOEHRD). (KRF-2006-214-D00064)

## Figure captions

**Figure 1.** Schematic view of metamaterial clad circular air hole waveguide structure and coordinates.

**Figure 2.** Dielectric and magnetic constants of metamaterial using in present study.

**Figure 3.** Three possible cases of allowed and forbidden regions for normalized propagation constant. Dashed region is forbidden. DNG or DPS clad cases of (a) $\mu_{r2}\varepsilon_{r2} > \mu_{r1}\varepsilon_{r1}\left(=1.0\right)$ and (b) $\mu_{r1}\varepsilon_{r1}\left(=1.0\right) > \mu_{r2}\varepsilon_{r2}$, and (c) ENG clad case.

**Figure 4.** Schematic drawing of field intensity of metamaterial clad circular air hole waveguide. (a) Surface plasmon polariton (SPP) mode and (b) circular waveguide (CWG) mode.

**Figure 5.** Forbidden and allowed regions of guided mode. Dashed region is forbidden.

**Figure 6.** Dispersion curves of TM-like principal modes for metamaterial clad circular air hole waveguide. (a) $TM_{01}$ mode, (b) $HE_{11}$ mode, and (c) $HE_{21}$ mode.

**Figure 7.** Dispersion curves of TM-like principal modes in case of $a = 15$ mm and $a = 1$ mm.

**Figure 8.** Dispersion curves of TE-like principal modes for metamaterial clad circular air hole waveguide. (a) $TE_{01}$ mode, (b) $EH_{11}$ mode, and (c) $EH_{21}$ mode.

**Figure 9.** Dispersion curves of TE-like principal modes in case of $a = 10$ mm and $a = 1$ mm.

**Figure 10.** Single mode backward wave region of $TM_{01}$ and $TE_{01}$ modes.

Figure 1

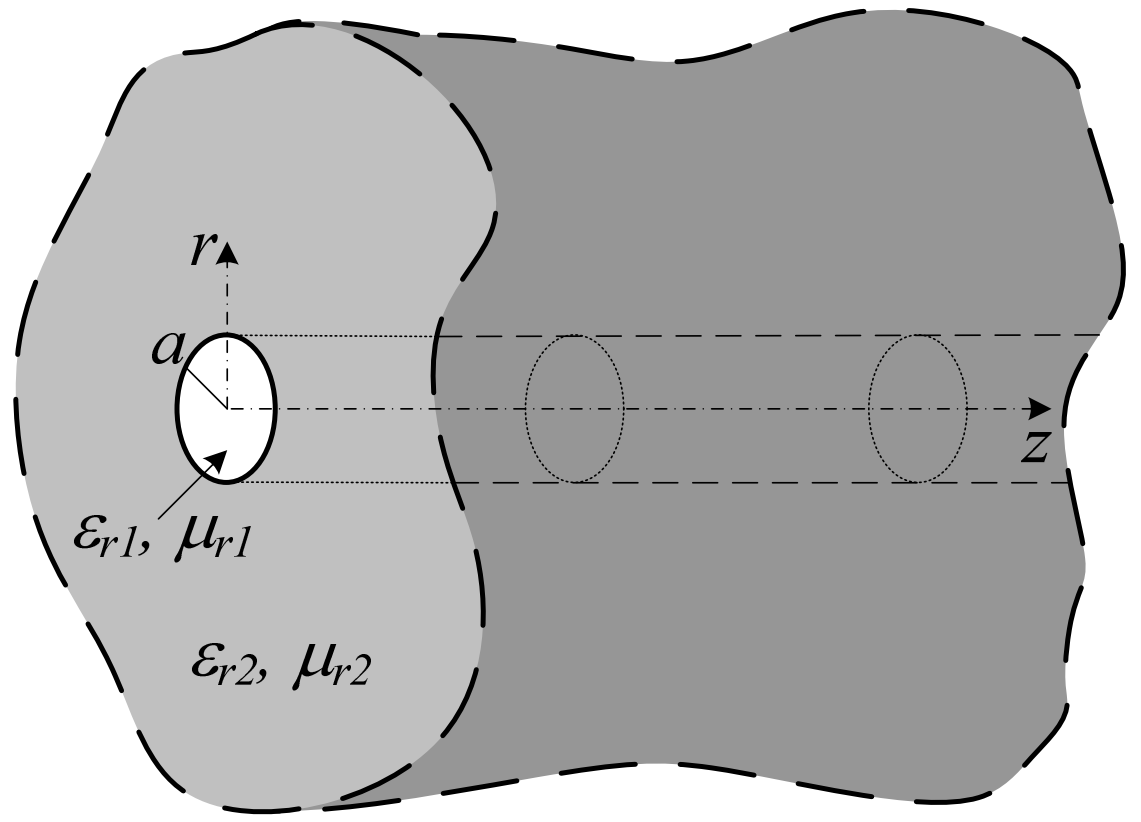

Figure 2

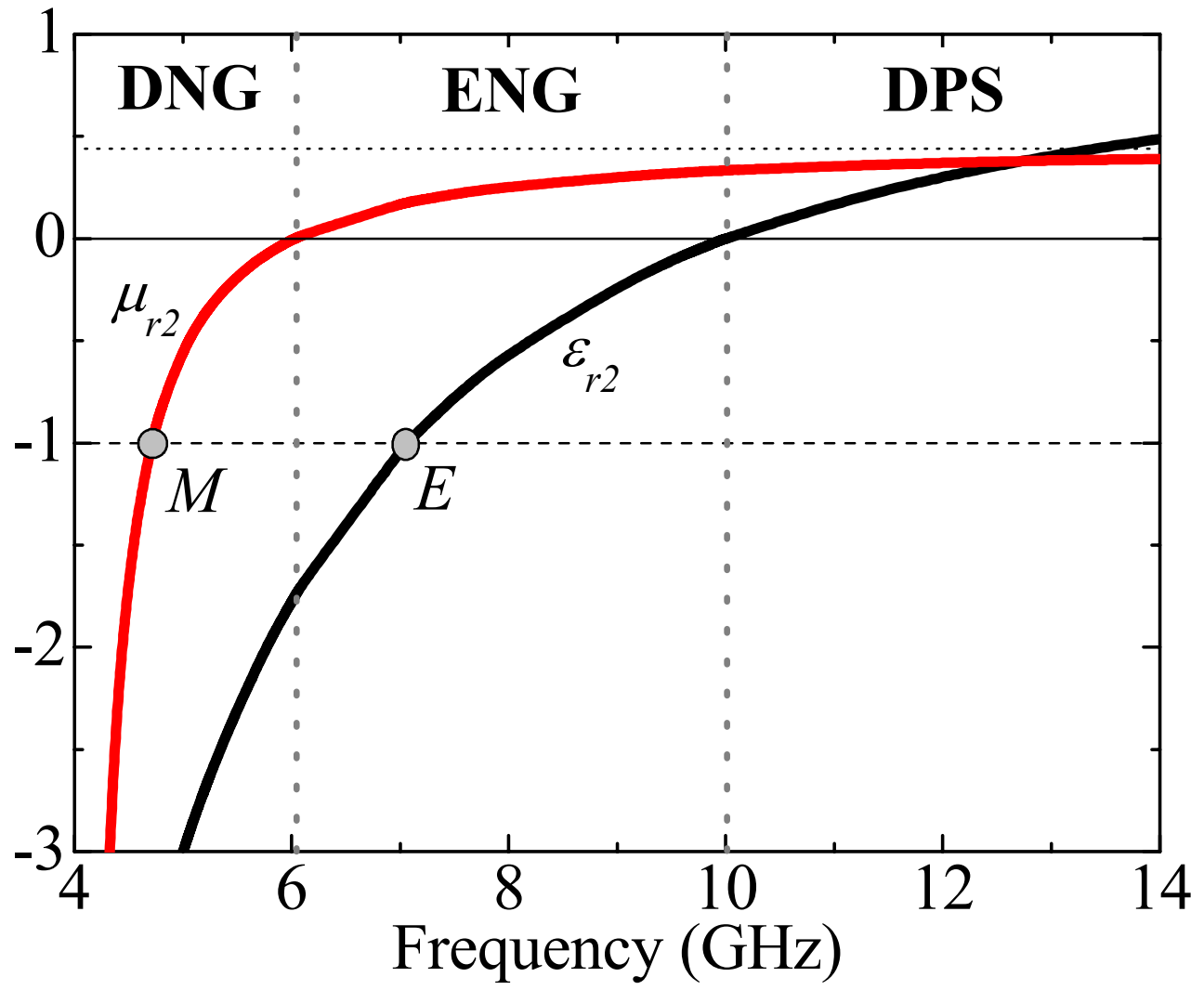

Figure 3

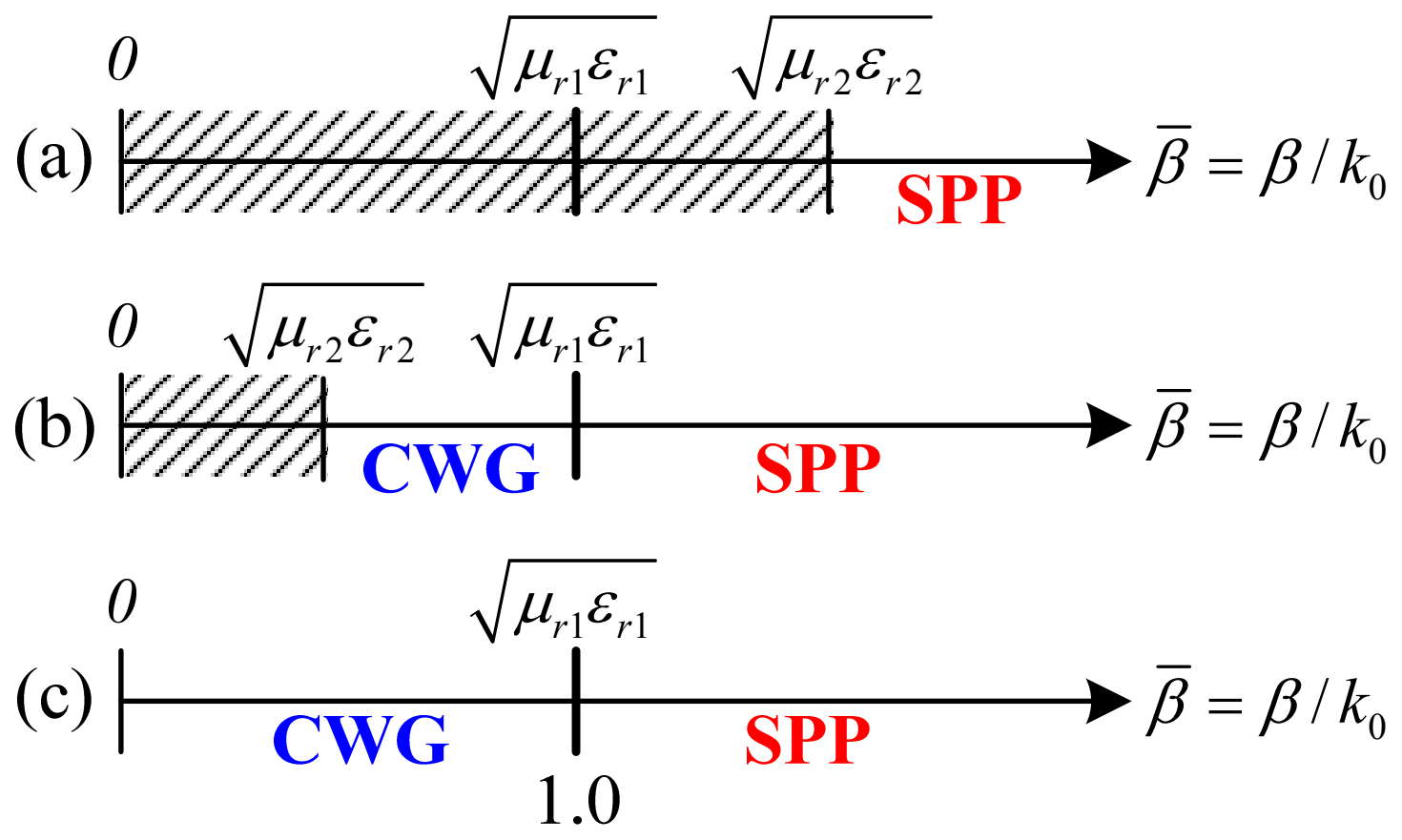

Figure 4

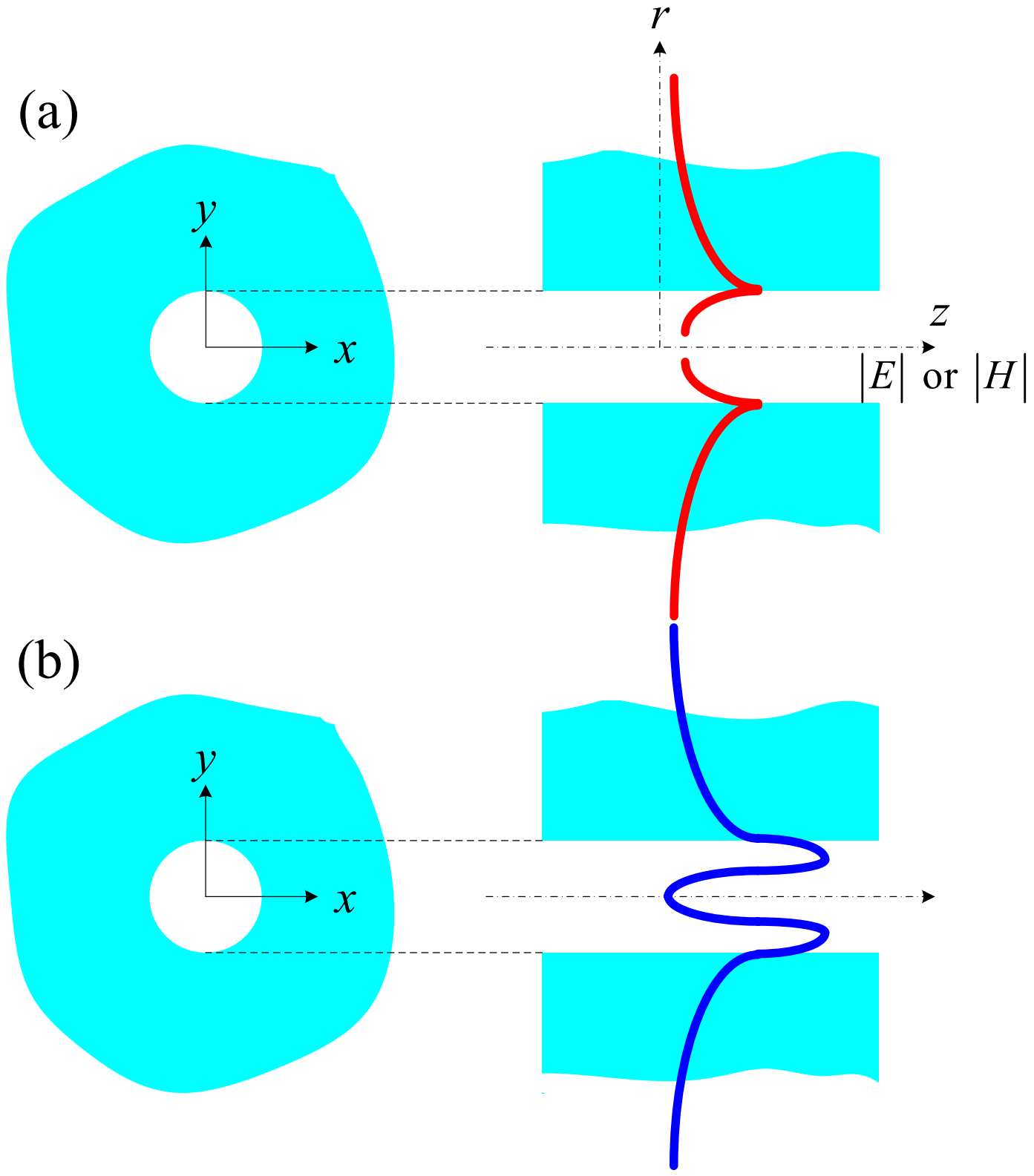

Figure 5

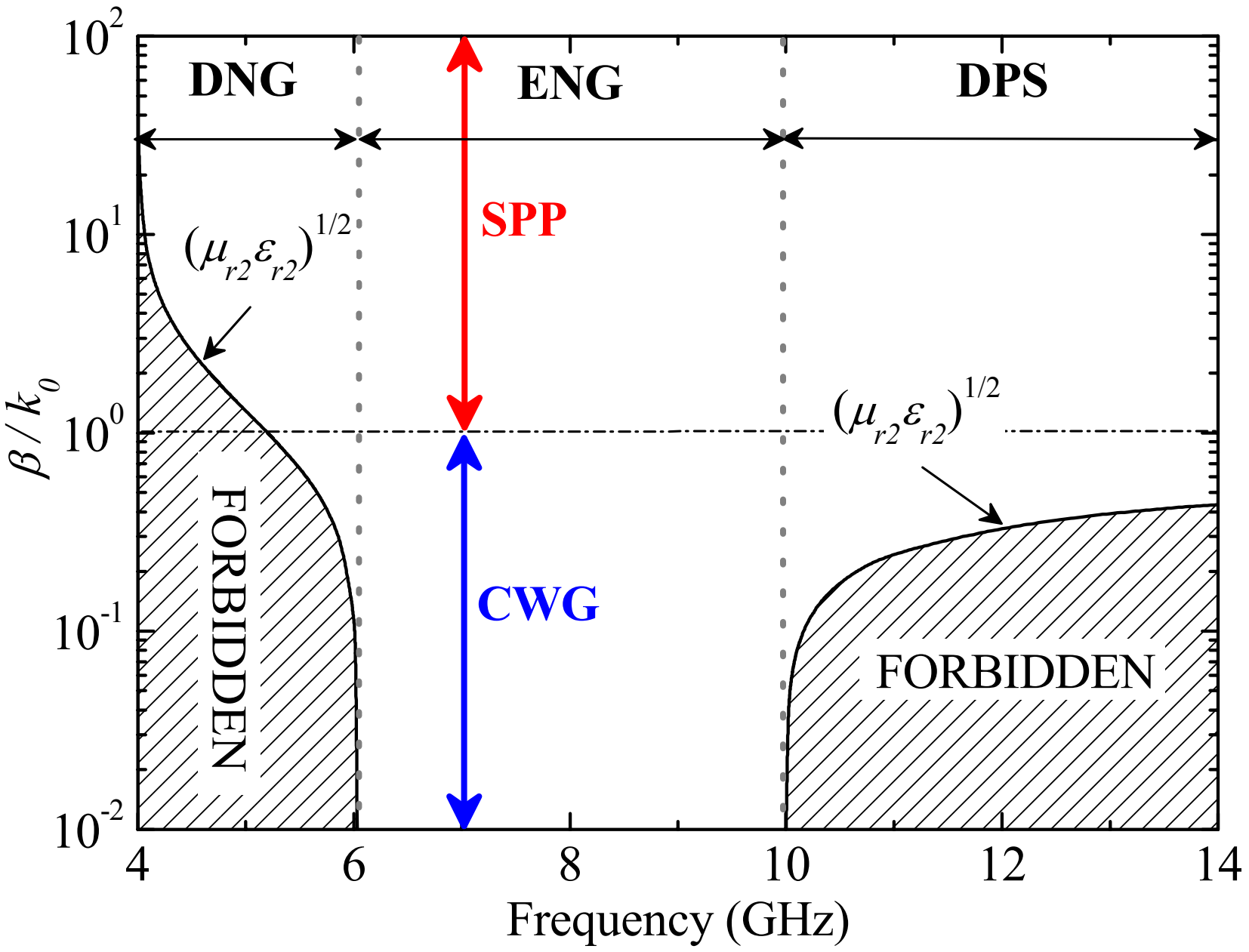

Figure 6

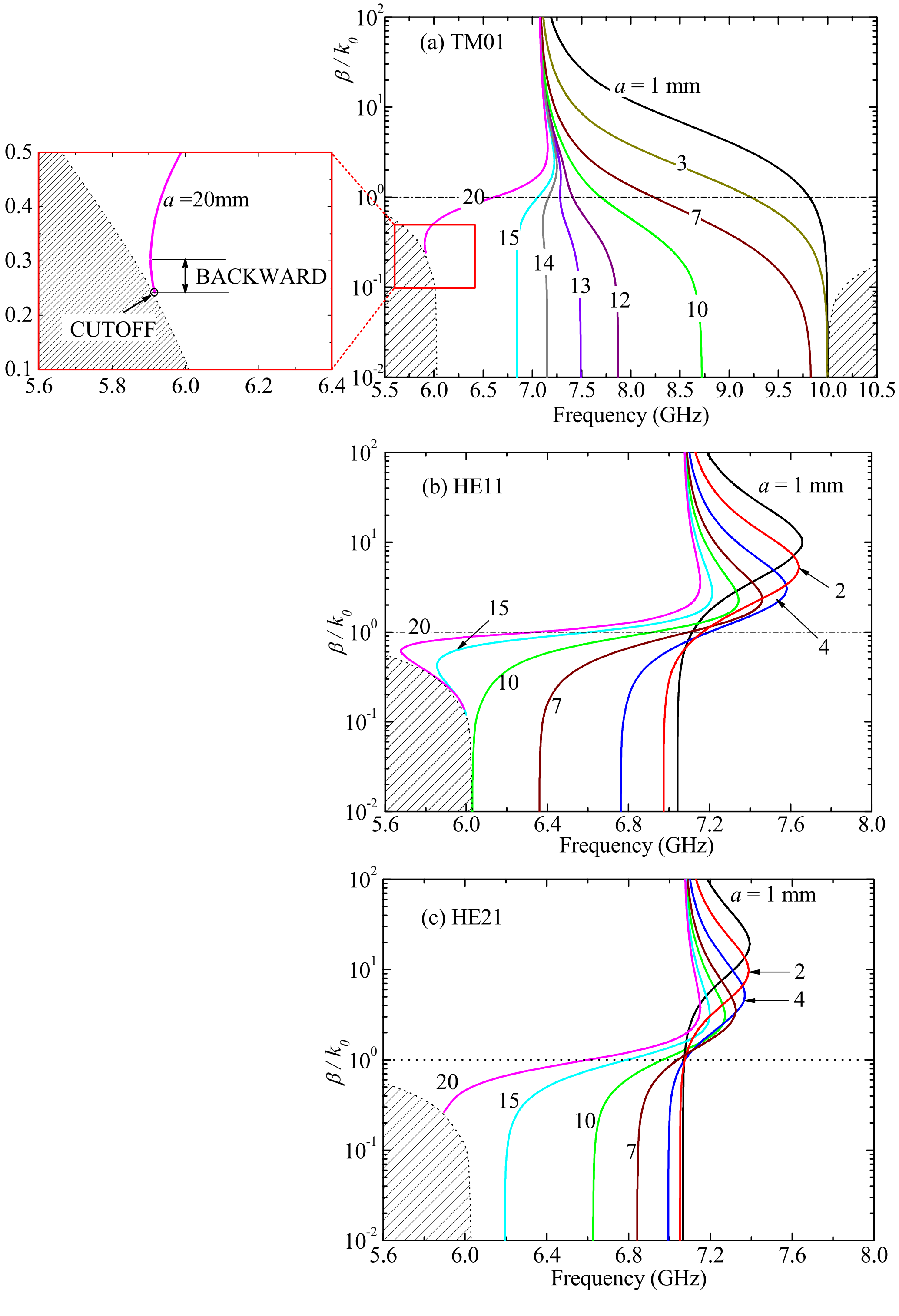

Figure 7

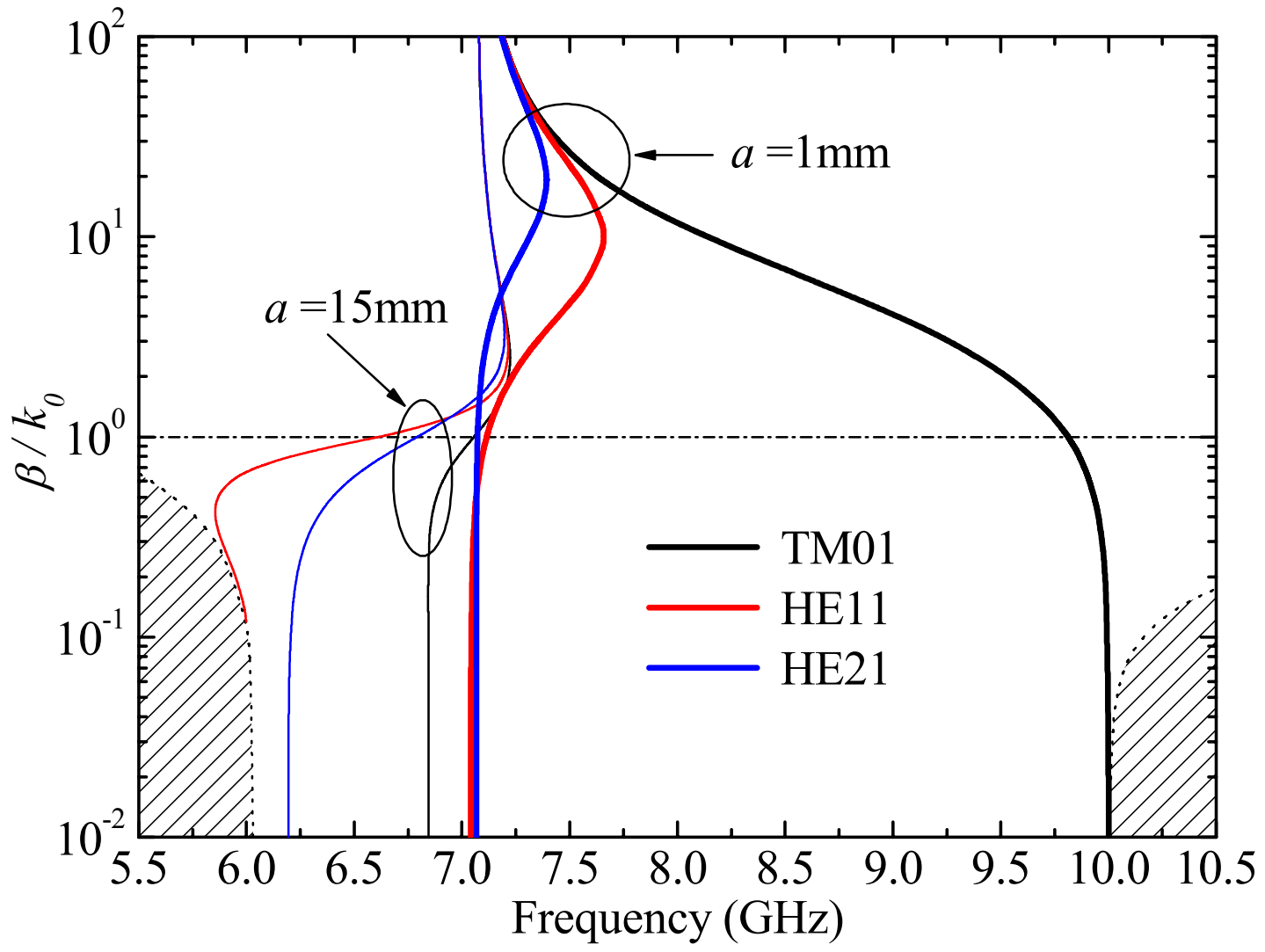

$10^2$
$10^1$
$10^0$
$10^{-1}$
$10^{-2}$
$\beta / k_0$
$a$ =1mm
$a$ =15mm
TM01
HE11
HE21
5.5 6.0 6.5 7.0 7.5 8.0 8.5 9.0 9.5 10.0 10.5
Frequency (GHz)

Figure 8

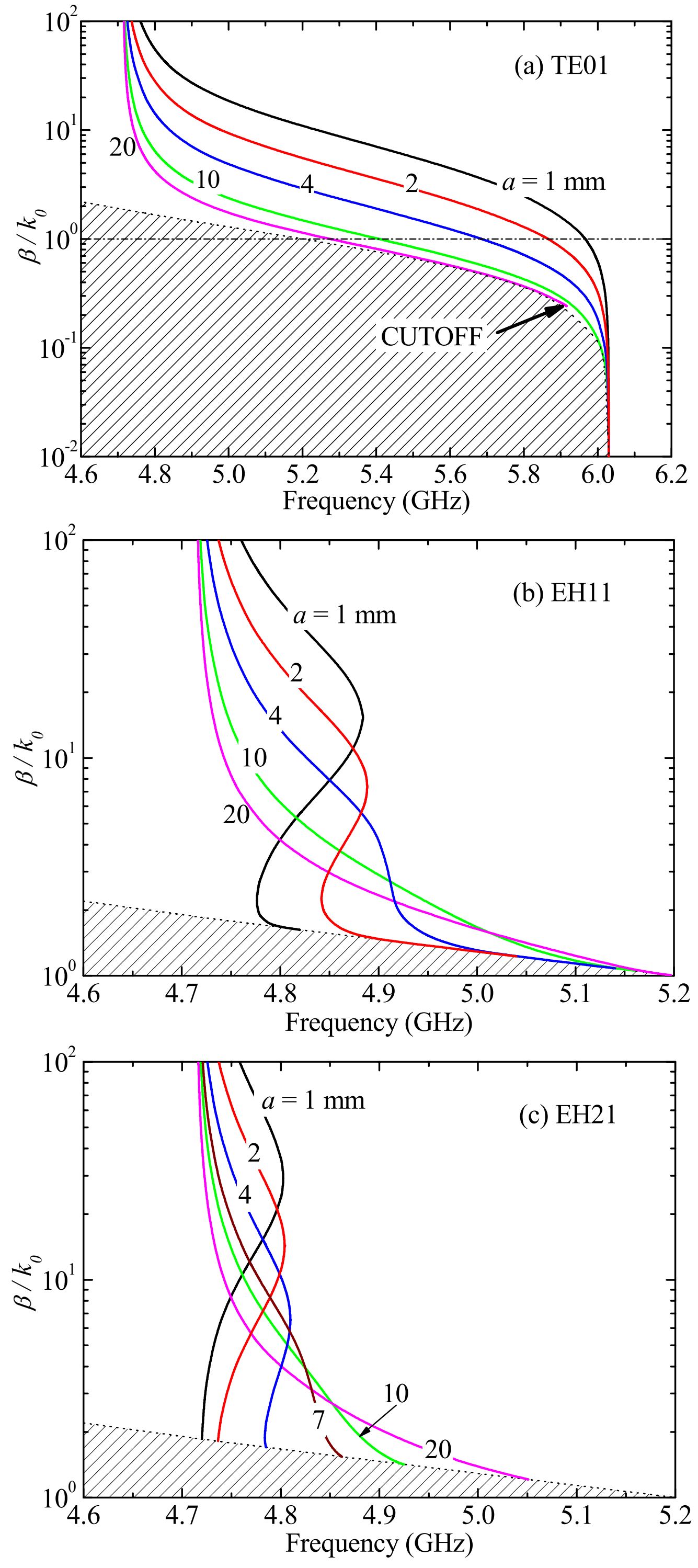

Figure 9

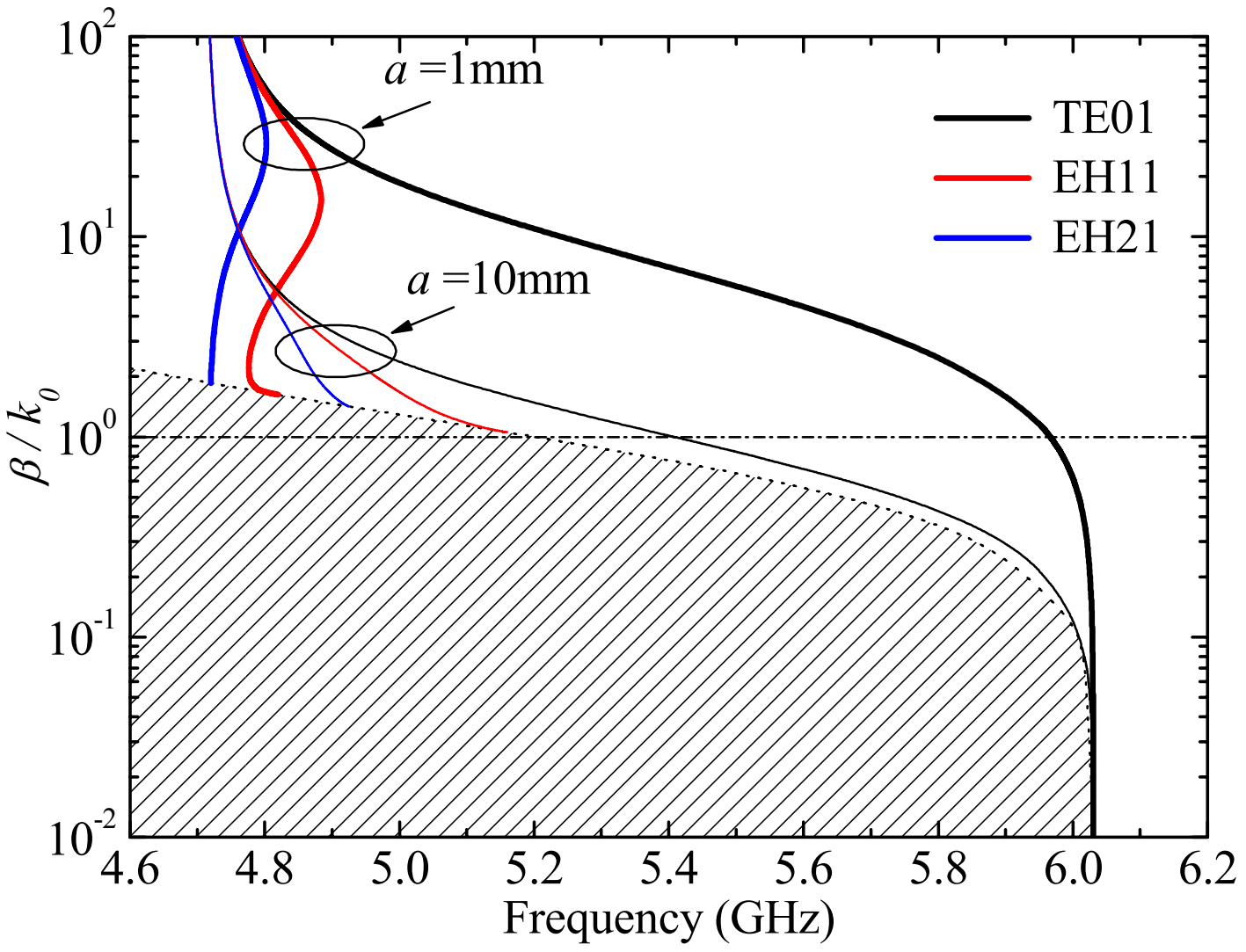

Figure 10

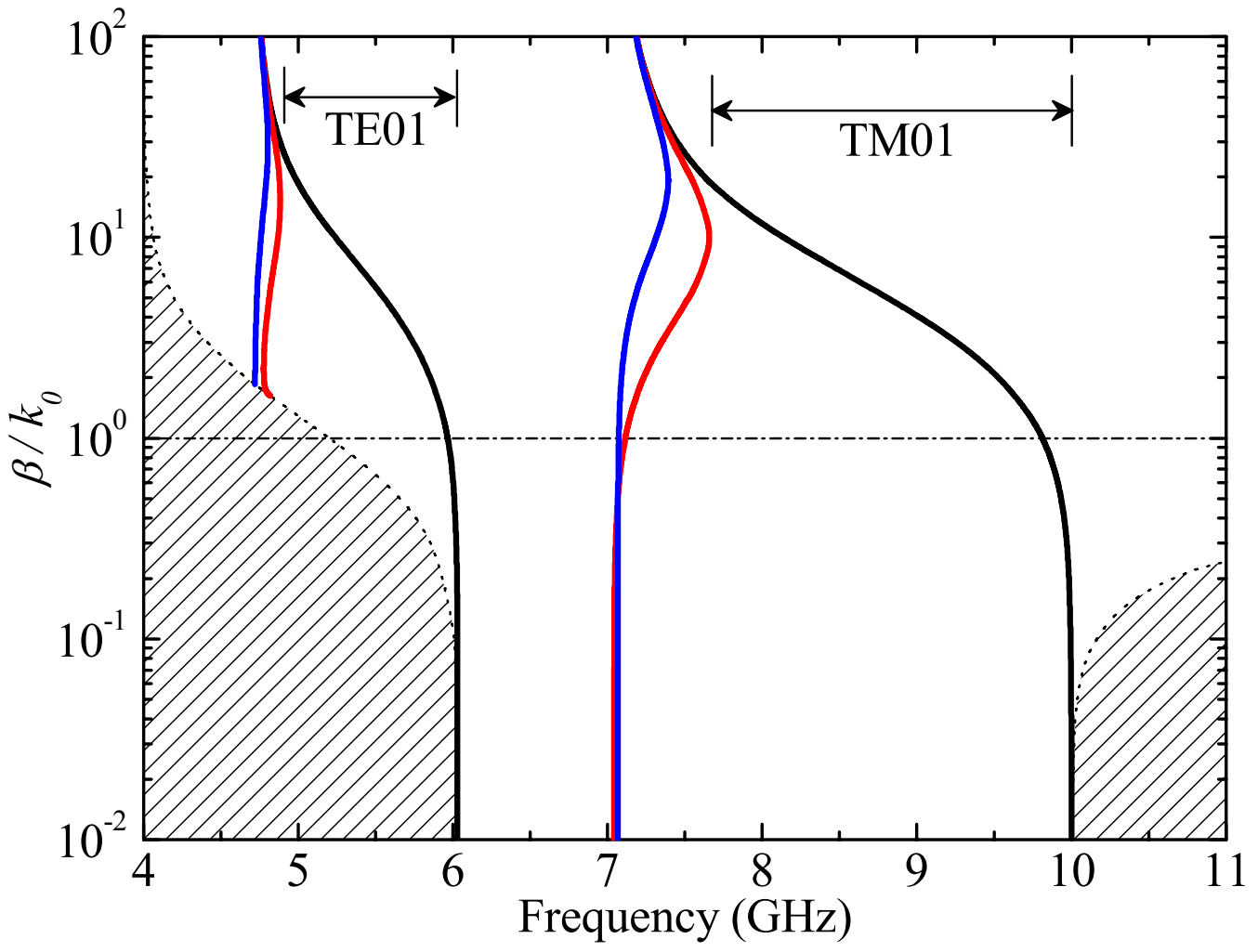